\documentstyle[amssymb,prl,aps,epsf,epsfig,floats,twocolumn]{revtex}
\begin{document}
\draft

\twocolumn[\hsize\textwidth\columnwidth\hsize\csname
@twocolumnfalse\endcsname

\title{
\hfill{\small{DOE/ER/40762-234}}\\
\hfill{\small{UMD-PP\# 02-001}}\\[0.6cm]
Constructing Parton Convolution in Effective Field Theory}
\author{Jiunn-Wei Chen and Xiangdong Ji}
\address{Department of Physics, University of Maryland,
College Park, MD 20742-4111}
\maketitle
\begin{abstract}
Parton convolution models have been used extensively
in describing the sea quarks in the nucleon and explaining
quark distributions in nuclei (the EMC effect). 
From effective field theory point of view, we construct 
the parton convolution formalism which has been the
underlying conception of all convolution models. 
We explain the significance of scheme and scale
dependence of the auxiliary quantities such as 
the pion distributions in a nucleon.
As an application, we calculate the complete leading 
nonanalytic chiral contribution to the isovector 
component of the nucleon sea. 

\end{abstract}
\medskip]

Calculating parton distributions from the first principles have been proven
difficult. The only approach at present is lattice field theory in which the
moments of parton distributions are simulated on a Euclidean lattice \cite
{MIT,qcdsf,sesam,liu}. A useful phenomenological approach used to understand
certain aspects of the parton distributions, such as their modifications in
a nucleus or the origin of sea quarks, is to construct a parton convolution
model \cite{sullivan,thomasx,others,review}. The essence of the parton
convolution picture is easy to describe: First, the system under
consideration is treated as a composite of hadrons (nucleons and their
excitations and mesons). Then hard scattering is pictured to happen on one
of its hadron constituents in which the parton distributions are presumed
known. Although widely used, the convolution models seem hard to justify at
a more fundamental level. For instance, it has been difficult to construct
systematic corrections to the model predictions.

In recent publications \cite{CJ,AS}, the effective field theory
techniques---more specifically chiral perturbation theory \cite{HBChPT,BKM}%
---were first used to understand the chiral corrections to the parton
distributions in the nucleon. We find that the approach can easily be
generalized to understand the sea quark distributions in the nucleon and the
modifications of parton distributions in nuclei. The result is a rigorous
and general way to construct parton distributions of composite systems out
of their hadron constituents, generalizing the conventional convolution
model approach. Unlike the convolution models, however, effective field
theories allow a systematic way to account for higher-order contributions
through power counting.

Our observation is simple: The moments of the parton distributions are
defined from the matrix elements of the twist-two operators. For instance,
for the unpolarized parton distributions these operators are 
\begin{equation}
{\cal O}_{q}^{\mu _{1}\cdots \mu _{n}}=\overline{q}\gamma ^{(\mu
_{1}}iD^{\mu _{2}}\cdots iD^{\mu _{n})}q\ .
\end{equation}
In effective field theories, these operators are matched to hadronic
operators with the same quantum numbers, 
\begin{equation}
{\cal O}_{q}^{\mu _{1}\cdots \mu _{n}}=\sum_{j=1}^{\infty }c_{qj}^{(n)}{\cal %
O}_{j}^{\mu _{1}\cdots \mu _{n}}\ ,  \label{expan}
\end{equation}
where $j$ labels different types of hadronic operators and $c_{qj}^{(n)}$
are c-number coefficients outside of the effective field theory framework.
As a specific example, the isovector operator 
\begin{eqnarray}
{\cal O}_{u-d}^{\mu _{1}\cdots \mu _{n}} &=&\bar{u}\gamma ^{(\mu
_{1}}iD^{\mu _{2}}\cdots iD^{\mu _{n})}u  \nonumber \\
&&-\bar{d}\gamma ^{(\mu _{1}}iD^{\mu _{2}}\cdots iD^{\mu _{n})}d\ ,
\end{eqnarray}
can be matched onto the hadronic operators 
\begin{eqnarray}
{\cal O}_{u-d}^{\mu _{1}\cdots \mu _{n}} &=&a_{n}\frac{f_{\pi }^{2}}{4}%
\left\{ {\rm Tr}\left[ \Sigma ^{\dagger }\tau _{3}iD^{(\mu _{1}}\cdots
iD^{\mu _{n})}\Sigma \right] \right.  \nonumber \\
&&\left. +{\rm Tr}\left[ \Sigma \tau _{3}iD^{(\mu _{1}}\cdots iD^{\mu
_{n})}\Sigma ^{\dagger }\right] \right\}  \nonumber \\
&&+b_{n}\overline{N}v^{(\mu _{1}}\cdots v^{\mu _{n})}\left( u\tau
_{3}u^{\dagger }+u^{\dagger }\tau _{3}u\right) N\ ,  \nonumber \\
&&+c_{n}\overline{N}S^{(\mu _{1}}v^{\mu _{2}}\cdots v^{\mu _{n})}\left(
u^{\dagger }\tau _{3}u-u\tau _{3}u^{\dagger }\right) N\   \nonumber \\
&&+\cdots \ ,  \label{3}
\end{eqnarray}
where $N$ and ${\overline{N}}$ are the nucleon fields in the nonlinear
representation, $v^{\mu }$ is the nucleon four-velocity, $u=\exp (i\pi
^{a}\tau ^{a}/2f_{\pi })$ with pion fields $\pi ^{a}$ and decay constant $%
f_{\pi }=93$ MeV, $\Sigma =u^{2},$ and $\left( \cdots \right) $ denotes the
symmetrization of the indices in between. We have neglected terms with more
derivatives as well as with multiple nucleon fields relevant for the nuclear
modifications of parton distributions.

Of course, the expressions in Eqs. (\ref{expan}) and (\ref{3}) are useless
if there are no small expansion parameters and every term contributes at
equal importance. If, however, a sensible expansion scheme exists, as when
an effective field theory becomes applicable, one can learn some aspects of
the parton distributions through the expansion. In particular, the expansion
allows the quark distributions in a hadron system $H$ be expressed as
convolutions; 
\begin{eqnarray}
q_{H}(x) &=&\sum_{j}\left[ \int_{|x|}^{1}{\frac{dy}{y}}q_{j}(y)f_{jH}\left( {%
\frac{x}{y}}\right) \right.  \nonumber \\
&&\left. -\int_{-1}^{-|x|}{\frac{dy}{y}}q_{j}(y)f_{jH}\left( {\frac{x}{y}}%
\right) \right] \ ,  \label{result}
\end{eqnarray}
where $q_{j}(y)$ is defined through its moments 
\begin{equation}
\int_{-1}^{+1}dyy^{n-1}q_{j}(y)=c_{qj}^{(n)}\ ,  \label{6}
\end{equation}
and can be interpreted as the quark distribution in a hadron state $j$. $%
f_{jH}(y)$ is defined through moments 
\begin{equation}
\int_{-1}^{+1}dyy^{n-1}f_{jH}(y)=\frac{1}{2(P^{+})^{n}}\langle P|{\cal O}%
_{j}^{+\cdots +}|P\rangle _{H}\ ,  \label{8}
\end{equation}
which can be interpreted as the hadron $j$ distribution in $H$ ($%
P^{+}=(P^{0}+P^{3})/\sqrt{2}$, and similarly for other indices). Using Eqs. (%
\ref{result})-(\ref{8}), it is easy to check that the $(n-1)$th moment of $%
q_{H}(x)$ reproduces Eq. (\ref{expan}). A similar convolution formula can be
derived for polarized distributions. Equation (\ref{result}) is schematic
because $j$ refers to general hadronic states having the quantum numbers of
the twist-two operators. For example, $j$ can be an $N-\Delta $ interference
state, or a 3-pion state. For intermediate states with multiple hadrons, $y$
must be extended to several light-cone variables. For off-diagonal hadron
states, the quark distributions are the generalized parton distributions
that have been studied recently in the literature \cite{ofpd}. According to
Eq. (\ref{result}), various convolution models consist of particular
truncations of the expansion.

To illustrate the above formalism, we study in the remainder of the paper
the moments of the proton's isovector anti-quark distribution, $\bar{u}_P(x)-%
\bar{d}_P(x)$, in chiral perturbation theory. A chiral contribution to this
quantity was first computed in Ref. \cite{thomas}, and our result partially
confirms the answer there. However, we have also found an additional
contribution which depends on the quark distribution in the chiral limit.

We start by expressing the isovector, spin-averaged twist-two quark
operators in terms of the operators with pure fields, 
\begin{eqnarray}
{\cal O}_{u-d}^{\mu _{1}\cdots \mu _{n}} &=&a_{n}\frac{f_{\pi }^{2}}{4}%
\left\{ {\rm Tr}\left[ \Sigma ^{\dagger }\tau _{3}iD^{(\mu _{1}}\cdots
iD^{\mu _{n})}\Sigma \right] \right.  \nonumber \\
&&\left. +{\rm Tr}\left[ \Sigma \tau _{3}iD^{(\mu _{1}}\cdots iD^{\mu
_{n})}\Sigma ^{\dagger }\right] \right\} +...  \nonumber \\
&=&-ia_{n}\epsilon ^{3ij}\pi ^{i}i\partial ^{(\mu _{1}}...i\partial ^{\mu
_{n})}\pi ^{j}+...\ ,  \label{7}
\end{eqnarray}
where after the second equal sign, we have made the chiral expansion and
neglected the higher-order terms in $m_{\pi }/(4\pi f_{\pi })$. The matrix
elements of the isovector operators, $O_{\pi ^{+}-\pi ^{-}}^{\mu _{1}\cdots
\mu _{n}}=-i\epsilon ^{3ij}\pi ^{i}i\partial ^{\mu _{1}}\cdots i\partial
^{\mu _{n}}\pi ^{j}$, in the proton state define the isovector pion
distribution $f_{(\pi ^{+}-\pi ^{-})/P}(y)$: 
\begin{eqnarray}
\langle P|O_{\pi ^{+}-\pi ^{-}}^{\mu _{1}\cdots \mu _{n}}|P\rangle
&=&2A_{n}P^{\mu _{1}}\cdots P^{\mu _{n}}\ ,  \nonumber \\
\int_{-1}^{+1}f_{({\pi ^{+}}-{\pi ^{-}})/P}(y)y^{n-1}dy &=&A_{n}\ .
\end{eqnarray}
The matrix elements $A_{n}$ can be calculated in chiral perturbation theory.
In dimensional regularization and heavy-nucleon formalism \cite{HBChPT}, we
find 
\begin{eqnarray}
A_{n\text{ odd }} &=&\left( -1\right) ^{\frac{n-1}{2}}\left( \frac{n+3}{n+1}%
\right) \frac{\left( 3g_{A}^{2}+\delta _{n1}\right) m_{N}^{2}}{4\left( 4\pi
f_{\pi }\right) ^{2}}  \nonumber \\
&&\times \left( \frac{m_{\pi }}{m_{N}}\right) ^{n+1}\log \frac{m_{\pi }^{2}}{%
m_{N}^{2}}+...\ ,  \nonumber \\
A_{n\text{ even}} &=&0\ ,  \label{an}
\end{eqnarray}
where $g_{A}$ is the neutron decay constant in the chiral limit. The
explicit nucleon mass dependence comes from the kinematic factors in the
definition of the matrix elements and from setting the renormalization scale
(in the chiral logarithm) to $m_{N}$. The ellipses denote terms analytic in
quark masses or subleading in chiral power counting. The $n\ =1$ moment is
special because it receives a pion tadpole contribution. The Bose symmetry
leads to vanishing $n=$ even moments.

From the moments, one can construct the pion distribution, $f_{(\pi ^{+}-\pi
^{-})/P}(y)$ in the proton, 
\begin{eqnarray}
f_{(\pi ^{+}-\pi ^{-})/P}(y) &=&\frac{m_{N}^{2}}{2\left( 4\pi f_{\pi
}\right) ^{2}}\left( \frac{m_{\pi }}{m_{N}}\right) ^{2}\log \frac{m_{\pi
}^{2}}{m_{N}^{2}}\delta (y)  \nonumber \\
&&+g(y)\theta (y)+g(-y)\theta (-y)\ ,  \label{fpi}
\end{eqnarray}
where 
\begin{eqnarray}
g(y) &=&\frac{-3g_{A}^{2}m_{N}^{2}}{4\left( 4\pi f_{\pi }\right) ^{2}}%
y\left\{ \frac{m_{\pi }^{2}\left( 1-y\right) }{m_{\pi }^{2}\left( 1-y\right)
+m_{N}^{2}y^{2}}\right.  \nonumber \\
&&+\left. \log \left( \frac{m_{\pi }^{2}\left( 1-y\right) +y^{2}m_{N}^{2}}{%
y^{2}m_{N}^{2}}\right) \right\} \ .
\end{eqnarray}
Apart from the leading nonanalytic contribution from small $y$, the above
expression also contains analytic terms in quark masses. The Bose symmetry
dictates $f_{(\pi ^{+}-\pi ^{-})/P}(y)$ as an even function of $y$. Equation
(\ref{fpi}) can also be derived from the definition of the pion distribution
as the matrix elements of the light-cone string operator in the same scheme 
\cite{collins} 
\begin{eqnarray}
f_{(\pi ^{+}-\pi ^{-})/P} &&(y)={\frac{1}{2P^{+}}}(-i\epsilon ^{3ij}) 
\nonumber \\
&&\times \int_{-\infty }^{\infty }{\frac{d\lambda }{2\pi }}e^{i\lambda
y}\langle P|\pi ^{i}(0)i\partial ^{+}\pi ^{j}(\lambda n)|P\rangle \ ,
\end{eqnarray}
where $n\sim (1,0,0,-1)$ is a light-cone vector. The $\delta $-function
contribution at $y=0$ comes from the pion-nucleon seagull vertex.

Before going further, it is useful to discuss the scheme and scale
dependence of the pion distribution in the nucleon. For simplicity, we match
a generic quark operator $O_{q}$ schematically onto the sum of a pion
operator $O_{\pi }$ and a nucleon operator $O_{N}$, 
\begin{equation}
O_{q}(\mu )=a(\mu ,\Lambda )O_{\pi }(\Lambda )+b(\mu ,\Lambda )O_{N}(\Lambda
) \ .  \label{14}
\end{equation}
The quark operator must be renormalized in QCD and has a perturbative QCD
renormalization scale $\mu $ much larger than the QCD scale, $\Lambda_{{\rm %
QCD}}$. The $\mu $ dependence in the right-hand side of the equation is
confined to the coefficient functions $a(\mu ,\Lambda )$ and $b(\mu ,\Lambda
)$. The hadronic operators, on the other hand, have an implicit scheme and
an explicit renormalization scale $\Lambda $ ($\lesssim 1$ GeV) dependence
in chiral perturbation theory. The scheme and scale dependence in these
operators ought be cancelled entirely by the scheme and scale dependence of
the coefficient functions. When Eq. (\ref{14}) is sandwiched in the nucleon
state, $\langle N|O_{\pi }|N\rangle $ is interpreted as the pionic effect in
the nucleon. Because of the scheme and scale dependence of the matrix
element, we cannot define in an absolute sense, for instance, the number of
pions in the nucleon.

In a typical calculation of the Sullivan process, the baryons (nucleons) are
treated fully relativistically with finite masses (see, for example, Ref. 
\cite{thomas}). The leading-order contribution is proportional to the
coupling 
\begin{equation}
\left( \frac{g_{\pi NN}}{4\pi }\right) ^{2}=\left( \frac{g_{A}m_{N}}{4\pi
f_{\pi }}\right) ^{2} \ ,
\end{equation}
where we have made use of the Goldberger-Treiman relation. Since the nucleon
mass $m_{N}$ is on the order of $4\pi f_{\pi }$, the expansion parameter is
by no means small. In particular, the multi-loop contributions are
proportional to $(g_{\pi NN}/4\pi )^{2n}$ which is not suppressed relative
to the leading order. Therefore, in this scheme of chiral expansion, $%
\langle N|O_{\pi }|N\rangle $ cannot be calculated with a controlled
approximation. As such, the leading order result can only be taken as a
model prediction even when the loop momentum of the pion is kept small. The
heavy-baryon chiral perturbation \cite{HBChPT} is designed to overcome this
difficulty. In this scheme, the leading-order result is quadratically
divergent. In a cut-off regularization, it is proportional to $%
g^2_A\Lambda^2 /(4\pi f_\pi)^2$, which is a small parameter when $\Lambda
\ll 4\pi f_\pi$. As a consequence, the pionic contribution is very sensitive
to the cut-off. In dimensional regularization, on the other hand, the power
divergences are absent by definition. From dimensional analysis, the matrix
element $\langle N|O_{\pi }|N\rangle $, and here the pion distributions in
the nucleon, vanishes in the chiral limit. The renormalization scale $%
\Lambda $ enters in the logarithms of the leading nonanalytical term and
controls the relative contribution to $O_q$ from $O_{\pi }$ and $O_{N}$.

The coefficient $a_{n}$ in Eqs. (\ref{3}) and (\ref{7}) is related to the $%
u-d$ distribution in the $\pi ^{+}$ meson: 
\begin{equation}
a_{n}=\int_{-1}^{+1}x^{n-1}(u_{\pi ^{+}}^{0}(x)-d_{\pi ^{+}}^{0}(x))dx\ ,
\end{equation}
where the superscript $0$ indicates the chiral limit. The isospin charge
counting gives $a_{1}=2$. The difference between $u_{\pi ^{+}}^{0}$ and the
real-world $u_{\pi ^{+}}$ is a chiral correction which has been computed in
Ref. \cite{AS}: 
\begin{eqnarray}
&u_{\pi ^{+}}&(x)-d_{\pi ^{+}}(x)  \nonumber \\
&=&\left( u_{\pi ^{+}}^{0}(x)-d_{\pi ^{+}}^{0}(x)\right) \times \left( 1-%
\frac{m_{\pi }^{2}}{\left( 4\pi f_{\pi }\right) ^{2}}\log \frac{m_{\pi }^{2}%
}{m_{N}^{2}}\right)  \nonumber \\
&&+\frac{2m_{\pi }^{2}}{\left( 4\pi f_{\pi }\right) ^{2}}\log \frac{m_{\pi
}^{2}}{m_{N}^{2}}\delta (x)\ ,
\end{eqnarray}
where the delta function contribution comes from isospin conservation. For
the following purpose, this correction is of higher order.

Through Eq. (\ref{7}), we construct a contribution to the proton's $%
u(x)-d(x) $ distribution from the pion cloud: 
\begin{eqnarray}
u_{\pi /P}(x) &&-d_{\pi /P}(x)=\int_{|x|}^{1}{\frac{dy}{y}}\Big\{u_{\pi
^{+}}^{0}(y)-d_{\pi ^{+}}^{0}(y)  \nonumber \\
&&-(\bar{u}_{\pi ^{+}}^{0}(y)-\bar{d}_{\pi ^{+}}^{0}(y))\Big\}f_{(\pi
^{+}-\pi ^{-})/P}\left( \frac{x}{y}\right) ,  \label{pi}
\end{eqnarray}
where we have used the symmetry property of $f_{(\pi ^{+}-\pi ^{-})/N}(y)$
in $y$ and the definition of antiquark distribution $\overline{q}%
_{j}(y)=-q_{j}(-y)$. Since the contribution is an even function of $x$, we
can easily calculate the resulting up-down asymmetry in the sea, 
\begin{eqnarray}
&&\int_{0}^{1}({\bar{d}}_{\pi /P}(x)-{\bar{u}}_{\pi /P}(x))dx  \nonumber \\
&&=\frac{1}{2}\int_{-1}^{1}({u}_{\pi /P}(x)-{d}_{\pi /P}(x))dx  \nonumber \\
&&=\frac{1}{2}a_{1}A_{1}  \nonumber \\
&&={\frac{3g_{A}^{2}+1}{2(4\pi f_{\pi })^{2}}}m_{\pi }^{2}\log (\frac{m_{\pi
}^{2}}{m_{N}^{2}})\ +\cdots \ .
\end{eqnarray}
Above result agrees with that in Ref. \cite{thomas} in the limit $g_{A}=1$.
A number of comments can be made about this result. First, $\int_{0}^{1}dx(%
\bar{d}(x)-\bar{u}(x))$ is not an usual moment of the quark distribution and
cannot be calculated as the matrix element of a local operator in QCD or
effective field theories. We overcome this difficulty by converting the
matrix elements of the complete tower of twist-two operators into a
distribution and then integrating over $x$. The simple result follows from
that the pion contribution to the quark distribution in the nucleon is even
in $x$. Therefore, in a certain sense, our derivation provides a formal
justification for the approach used in Ref. \cite{thomas}. Second, the
difference between our result and that of Ref. \cite{thomas} comes from the
use of the linear sigma model in the later paper. As was discussed in Ref. 
\cite{CJ}, in a general formulation of chiral expansion in which the nucleon
fields furnish a linear representation of the chiral group, the difference
disappears. Finally, the above nonanalytic chiral contribution to $%
\int_{0}^{1}dx(\bar{d}(x)-\bar{u}(x))$ has an opposite sign from the
experimental data \cite{data}, indicating the importance of other
contributions to the nucleon sea.

According to the expansion in Eq. (\ref{3}), the twist-two pion operators
are not the only source of chiral logarithms in the antiquark distributions
in the nucleon. They can also be generated from the nucleon operators. Using 
$u_{P}^{0}(x)$ and $d_{P}^{0}(x)$ to represent the quark distributions in
the proton in the chiral limit, the result in Refs. \cite{CJ,AS} allows us
to find another chiral contribution to $u$ and $d$ distributions in the
proton, 
\begin{eqnarray}
&&u_{N/P}(x)-d_{N/P}(x)  \nonumber \\
&=&\left( u_{P}^{0}(x)-d_{P}^{0}(x)\right) \left( 1-\frac{\left(
3g_{A}^{2}+1\right) m_{\pi }^{2}}{\left( 4\pi f_{\pi }\right) ^{2}}\log 
\frac{m_{\pi }^{2}}{m_{N}^{2}}\right) \ .
\end{eqnarray}
Integrating from $-1$ to 0, we have 
\begin{eqnarray}
\int_{0}^{1}({\bar{d}}_{N/P}(x)- &&{\bar{u}}_{N/P}(x))dx=\int_{0}^{1}({\bar{d%
}^{0}}_{P}(x)-{\bar{u}^{0}}_{P}(x))dx  \nonumber \\
&&\times \left( 1-\frac{\left( 3g_{A}^{2}+1\right) m_{\pi }^{2}}{\left( 4\pi
f_{\pi }\right) ^{2}}\log \frac{m_{\pi }^{2}}{m_{N}^{2}}\right) \ .
\end{eqnarray}
The second term in the bracket represents a chiral contribution which cannot
be calculated without the knowledge of $\int_{0}^{1}({\bar{d}^{0}}_{P}(x)-{%
\bar{u}^{0}}_{P}(x))dx$. Together with the result in Eq. (\ref{pi}), we have
the complete leading chiral contribution to the isovector antiquark
distributions in the nucleon.

The leading chiral contribution to the polarized up and down quark
asymmetry in the proton sea can be calculated in a similar way. The
quark-helicity up-down asymmetry in the sea is
\begin{eqnarray}
\int_{0}^{1}(\Delta {\bar{d}}_{P}(x)-\Delta  &&{\bar{u}}_{P}(x))dx=%
\int_{0}^{1}(\Delta {\bar{d}^{0}}_{P}(x)-\Delta {\bar{u}^{0}}_{P}(x))dx 
\nonumber \\
&&\times \left( 1-\frac{\left( 2g_{A}^{2}+1\right) m_{\pi }^{2}}{\left( 4\pi
f_{\pi }\right) ^{2}}\log \frac{m_{\pi }^{2}}{m_{N}^{2}}\right) \ ,
\end{eqnarray}
where the superscript $0$ indicates the chiral limit. Similarly, the
quark-transversity up-down asymmetry in the sea is
\begin{eqnarray}
\int_{0}^{1}(\delta {\bar{d}}_{P}(x)- &&\delta {\bar{u}}_{P}(x))dx=%
\int_{0}^{1}(\delta {\bar{d}^{0}}_{P}(x)-\delta {\bar{u}^{0}}_{P}(x))dx 
\nonumber \\
&&\times \left( 1-\frac{\left( 4g_{A}^{2}+1\right) m_{\pi }^{2}}{2\left(
4\pi f_{\pi }\right) ^{2}}\log \frac{m_{\pi }^{2}}{m_{N}^{2}}\right) \ .
\end{eqnarray}
In both cases, the contribution from the pion cloud is of
higher order in chiral power counting.

\acknowledgements
We thank T. Cohen for useful discussions on the subject of the paper.
This work is supported in part by the U.S. Dept. of Energy under grant No.
DE-FG02-93ER-40762.

\end{document}